\documentclass[twocolumn,showpacs,preprintnumbers,amsmath,amssymb, superscriptaddress]{revtex4}

\usepackage{multirow}
\usepackage{amsthm}
\usepackage{amssymb}
\usepackage{amsfonts}
\usepackage{bbm}
\usepackage[dvips]{graphicx}
\usepackage{subfigure}

\newtheorem{theorem}{Theorem}

\newcommand{\Tr}{\operatorname{Tr}}

\newcommand{\beq}{\begin{equation}}
\newcommand{\eeq}{\end{equation}}
\newcommand{\idmat}{{\mathbbm{1}}}
\newcommand{\idmap}{{\rm id}}
\newcommand{\bra}[1]{\langle #1 |}
\newcommand{\ket}[1]{| #1 \rangle}

\newcommand{\pro}[1]{\ket{#1}\bra{#1}}

\begin{document}

\title{Class of PPT bound entangled states associated to almost any set of pure entangled states}
\author{Marco Piani}
\email{piani@ts.infn.it}
\affiliation{
Institute of Theoretical Physics and Astrophysics, University of Gda\'nsk, 80--952 Gda\'nsk, Poland}
\author{Caterina E. Mora}
\email{caterina.mora@uibk.ac.at}
\affiliation{Institut f\"ur Quantenoptik und Quanteninformation
  der \"Osterreichischen Akademie der Wissenschaften, Innsbruck,
  Austria}

\begin{abstract}
\noindent We analyze a class of entangled states for bipartite
$d\otimes d$ systems, with $d$ non-prime. The entanglement of such
states is revealed by the construction of canonically associated
entanglement witnesses. The structure of the states is very simple
and similar to the one of isotropic states: they are a mixture of
a separable and a pure entangled state whose supports are
orthogonal. Despite such simple structure, in an opportune
interval of the mixing parameter their entanglement is not
revealed by partial transposition nor by the realignment
criterion, i.e. by any permutational criterion in the bipartite
setting. In the range in which the states are Positive under
Partial Transposition (PPT), they are not distillable; on the
other hand, the states in the considered class are provably
distillable as soon as they are Nonpositive under Partial
Transposition (NPT). The states are associated to any set of more
than two pure states. The analysis is extended to the multipartite
setting. By an opportune selection of the set of multipartite pure
states, it is possible to construct mixed states which are PPT
with respect to any choice of bipartite cuts and nevertheless
exhibit genuine multipartite entanglement. Finally, we show that
every $k$-positive but not completely positive map is associated
to a family of nondecomposable maps.
\end{abstract}

\maketitle

\section{Introduction}
\label{sec:introduction}

Entanglement is a resource required in many task typical of the
fields of quantum information and quantum computation
\cite{NC,alber2001}, like quantum
teleportation~\cite{teleportation} and superdense
coding~\cite{BennettWiesner}. While there is a clear definition of
what an entangled state is \cite{Werner1989},  it is in general
difficult to determine whether a given state is entangled or not.
Correspondingly, the structure of the space of states, as
classified with respect to the entanglement property, is still a
central issue of investigation. Moreover, in the multipartite case
the picture is even more complicated, since it appears that there
are qualitatively different kinds of
entanglement~\cite{DurVC00-wstate}.

We remark that the study of very specific and simply parametrized
classes of states, typically satisfying some symmetry (like Werner
states~\cite{Werner1989} or isotropic states~\cite{reduction}),
has always turned out to be very useful to improve our
understanding of the entanglement phenomenon and of the geometry
and properties of the set of states. In this paper we provide
examples of states that, despite a simple structure, exhibit
interesting properties both in the bipartite and multipartite
setting.

One of the means to investigate the entanglement of states is
based on the use of linear maps ~\cite{Peres96,sep1996} which are
positive (P) but not completely positive (CP): we shall refer to
them as PnCP maps. A map is P if it transforms any state into
another positive operator. It is moreover CP if also its partial
action on a subsystem of any larger system gives rise to a P map.
In the case of a bipartite system, a state is entangled if and
only if there exists a PnCP map such that the operator obtained
acting with the map on only one of the two subsystems is not
positive any more. The simplest example of PnCP map is the
operation of transposition $T$ (with respect to a given basis).
The action of transposition on one of the subsystems is called
\emph{partial transposition} (PT) and is also known as the
Peres-Horodecki criterion ~\cite{Peres96,sep1996}. In the
bipartite $2\otimes2$ and $2\otimes3$ dimensional cases, PT can
``detect'' all entangled states: only states that develop negative
eigenvalues under PT (NPT states) are entangled. In higher
dimensions there are states which are positive under partial
transposition (PPT) even if entangled. The latter states have the
interesting property that their entanglement cannot be distilled
(see \cite{alber2001} for a review), thus it is considered to be
``bound''.

Beside partial transposition, there is another easily-computable
entanglement criterion,
realignment~\cite{Rudolph2002-criterion,Chen2002-criterion}. Both
PT and realignment are part of the larger family of permutational
criteria~\cite{HHHcontractions,permutations}, and they constitute
the only two independent criteria of such type in the bipartite
scenario. It must be remarked that realignment (i) is not related
to a positive linear map and (ii) can detect some PPT bound
entangled states.

If we want to use linear PnCP maps to detect PPT bound entangled
states, it is necessary to use PnCP maps that are not
decomposable, i.e. that cannot be written as the sum of a CP map
and a CP map composed with transposition. Indeed, the study of P
maps is strictly related to the study of entanglement, the link
being provided by the Choi-Jamio{\l}kowsky isomorphism
\cite{choi,Jamiolkowski}.

It was proved that every entangled state is useful for tasks that
it would be impossible to perform classically \cite{masanes}; in
this sense bound entanglement can be
``activated''~\cite{activation}. Quite interestingly, it was found
that PPT bound entangled states provide probabilistic
interconvertibility among multipartite pure states, which are not
interconvertible by Local Operations and Classical Communication
(LOCC) alone \cite{boundishizaka,IPppt}.

The first explicit examples of a PPT entangled states were given
in \cite{Pawel97}, and, since then, many other examples have been
found \cite{PawelLewenstein,
BrussPeres2000,YuLiu,piani2006,FeiLiSun}. It has been shown
\cite{ChruscKossa} that most of these states are part of a same
family of PPT -- but not a priori entangled -- states. A first
systematic method to construct PPT bound entangled states was
proposed in \cite{Bennett99}, and is based on the concept of
unextendible product basis.

In the present work we first consider a class of bipartite $d\otimes d$
states, with $d$ a non-prime dimension, which are
constructed/described by a given set of pure states and a mixing
parameter (a probability). Given two states $\psi^{(1)}$ and
$\psi^{(2)}$, in a $d_1\otimes d_1$ and $d_2 \otimes d_2$ Hilbert
space, respectively, we consider the set of mixed states
parametrized by the mixing parameter $p$
\begin{multline}
\label{eq:structstate}
\rho_p(\psi^{(1)},\psi^{(2)})=\\
\frac{1-p}{(d_1^2-1)(d_2^2-1)}
(\idmat-P_{\psi^{(1)}})\otimes(\idmat-P_{\psi^{(2)}})\\
+p
P_{\psi^{(1)}}\otimes P_{\psi^{(2)}},
\end{multline}
where
$P_{\psi^{(i)}}$ is the projector onto state $\psi^{(i)}$.
We will see that,
for almost any choice of $\psi^{(1)}$ and $\psi^{(2)}$ entangled,
there exists a $p_\Gamma>0$ such that $\rho_p$ is PPT entangled for all
choices of $p<p_\Gamma$. The structure of these states can be
considered very simple in comparison to the PPT bound entangled
states already known in literature. The class is a generalization
of states already appeared in~\cite{boundishizaka,IPppt}, where it
was proved that some states in the class, even though PPT, are
entangled, because they allow operations that are impossible by
LOCC. In our case, we prove that they are entangled by
constructing canonically associated entanglement witnesses.
Moreover, we show that, for almost all choices of the involved
pure states and for a suitable range of the mixing parameter, the
states are entangled but detected by neither partial transposition
nor, in the bipartite case, realignment. This class of states is
naturally rich. Furthermore, it can be verified by direct
inspection that it is not contained in the class described in
\cite{ChruscKossa}, thus is contributes effectively to the variety
of the known PPT (entangled) states.

We also study the multipartite setting, to which the family of
states can be naturally extended~\cite{boundishizaka,IPppt}. It is
possible to show~\cite{boundishizaka} that these states can be PPT
entangled with respect to every bipartite cut. As in the bipartite
case, this is proved by associating to each state a canonical
witness. Furthermore, we find conditions for which the states
contain genuine multipartite entanglement, and show that it is
possible to have a genuinely multipartite entangled state which is
PPT with respect to any bipartite cut. Finally, we relate the
properties of $k$-positivity and nondecomposability of linear maps
(to be defined in the following), and show that even a
decomposable map can become useful to detect PPT entangled states
just by considering its trivial extensions.

The paper is organized as follows. In Section II we provide
definitions and basic notions. In Section III  we introduce the
basic set of states of interest, involving the choice of two pure
states, and in Section IV we associate to them canonical
witnesses. In Section V we discuss partial transposition and
realignment and in Section VI we generalize the construction of
the class of states relating it to the choice of a set of $M\geq2$
pure states. Section VII is devoted to some considerations
regarding the construction of the canonical witnesses. In Section
VII the multipartite setting is studied. In Section IX, starting
from considering tensor-like witnesses, we provide a general
theorem relating the properties of $k$-positivity and
nondecomposability.

\section{Definitions and basic notions}

A $d$-dimensional system is associated to the Hilbert space
$\mathbb{C}^d$, and operators on such system are described by the
algebra of $d\times d$ matrices with complex entries $M_d$. A
state $\rho$ corresponds to a normalized ($\Tr(\rho)=1$) positive
semidefinite ($\rho\geq0$) matrix. We will denote (normalized)
vectors in the Hilbert space by $\ket{\psi}$ or $\psi$, and the
projector onto the pure state $\psi$ by $P_\psi=\pro{\psi}$.

\subsection{Entanglement and separable states}

A bipartite system AB is associated to a tensor-product Hilbert
space $\mathcal{H}_{AB}=\mathcal{H}_A\otimes\mathcal{H}_B$. A pure
bipartite state $\psi_{AB}$ is entangled if it is not factorized,
i.e. not of the form $\psi_{AB}=\psi_A\otimes\psi_B$. A bipartite
mixed state $\rho_{AB}$ is separable if it can be written as a
convex combination of factorized states \beq \label{eq:sepstates}
\rho_{AB}=\sum_ip_iP_{\psi^i_A}\otimes P_{\psi^i_B},\quad
p_i\geq0,\quad\sum_ip_i=1, \eeq otherwise it is entangled.

More in general one can consider $N$-partite systems, that are
associated to tensor-product Hilbert spaces of the form
$\bigotimes_{i=1}^N\mathcal{H}_i$, where $\mathcal{H}_i$ is the
Hilbert space associated to system $i$. In this case it is
possible to study the separability issue with respect to different
groupings of the parties. A pure $N$-partite state $\psi_N$ is
$k$-separable if it can be written as a tensor product of $k$
states, i.e. as $\psi_N=\bigotimes_{i=1}^k \psi_{S_i}$, with
$\mathcal{P}_k=\{S_i\}_{i=1}^k$ a partition of the parties in $k$
subsets. In particular, $\psi_N$ is biseparable if
$\psi_N=\psi_{S_1}\otimes\psi_{S_2}$. A pure state is $k$-partite
entangled if it cannot be written as the tensor product of states,
each of which pertains to less than $k$ parties. Similarly, a
mixed state is $k$-separable if it can be written as a convex
combination of $k$-separable pure states. The $k$-partition need
not be the same for all the $k$-separable pure states entering in
the convex combination; if all the pure states can be chosen to be
$k$-separable with respect to the same partition $\mathcal{P}_k$,
we say that the state is $k$-separable with respect to the
partition $\mathcal{P}_k$. In particular, we say that a state is
biseparable if it is $2$-separable, and that it is separable along
a cut $S_1:S_2$ if it is $2$-separable with respect to the
partition $\{S_1,S_2\}$. A mixed state is $k$-partite entangled if
every possible convex decomposition of the state contains at least
a $k$-partite entangled pure state. Notice that a $N$-partite
state is biseparable if and only if it is not $N$-partite
entangled. Any result valid in the bipartite setting can be
applied to the multipartite case when considering a given cut.

In the bipartite case, any pure state $\psi$ can be written in its standard Schmidt decomposition:
\[
\ket{\psi}=\sum_{i=1}^{r}\mu_i\ket{i_{A}\otimes i_{B}},
\]
where $\mu_i>0$, $\sum_{i=1}^r\mu_i^2=1$, are the Schmidt
coefficients, $r\leq\min(d_A,d_B)$ is the Schmidt rank (or number)
and $\ket{i_{A(B)}}$ are orthogonal states (i.e. they can be
extended to an orthonormal basis). We say that a bipartite density
matrix $\sigma$ has Schmidt number $k$ if (i) for any
decomposition $\{p_i\geq0,\phi_i\}$ of $\sigma$, i.e.
$\sigma=\sum_ip_i\pro{\phi_i}$, at least one of the vectors
$\phi_i$ has at least Schmidt rank $k$ and (ii) there exists a
decomposition of $\sigma$ with all vectors $\{\phi_i\}$ of Schmidt
rank at most $k$~\cite{Schmidtnumber}.

\subsection{Partial transposition and realignment}
\label{sec:PTrealigment}

We recall now the two separability criteria which we will use in
the following and which are based on the reordering of the entries
of the density matrix: partial
transposition~\cite{Peres96,sep1996} and
realignment~\cite{Rudolph2002-criterion,Chen2002-criterion}. Given
a bipartite density matrix
$\rho=\sum_{ijkl}\rho_{ij,kl}\ket{ij}\bra{kl}$ the linear
operations of partial transposition and realignment are defined as
follows. Partial transposition (with respect to the first system)
corresponds to the reordering
$(\ket{ij}\bra{kl})^{\Gamma_A}=\ket{kj}\bra{il}$, and realignment
to $R(\ket{ij}\bra{kl})=\ket{ik}\bra{jl}$. It is immediate to see
that, if a state is separable, then both
$\|\rho_{AB}^{\Gamma_A}\|_1\leq1$ and $\|R(\rho_{AB})\|_1\leq1$
must hold, with $\|X\|_1=\Tr\sqrt{X^\dagger X}$ the trace norm of
$X$. The condition $\|\rho_{AB}^{\Gamma_A}\|_1\leq1$ is equivalent
to requiring that $\rho_{AB}$ stays positive under partial
transposition, i.e. $\rho_{AB}^{\Gamma_A}\geq0$.

As regards partial transposition, we note that for any bipartite
state $\ket{\psi}=\sum_j\mu_j\ket{ii}$ (here written in its
Schmidt decomposition), we have
\begin{multline}
(\pro{\psi})^\Gamma=\sum_j\mu_j^2\pro{jj}\\
+\sum_{j>i}\mu_i\mu_j(\pro{\psi^+_{ij}}-\pro{\psi^-_{ij}}),
\end{multline}
with $\ket{\psi^\pm_{ij}}=(\ket{ij}\pm\ket{ji})/\sqrt{2}$, and
where partial transposition was operated in the Schmidt basis. The
eigenvalues of $(\pro{\psi})^\Gamma$ are $\lambda^i_0=\mu_i^2$,
for $i=1,\ldots,d$, and $\lambda^{ij}_\pm=\pm\mu_i\mu_j$, for
$j>i$, corresponding to Schmidt-rank-one eigenstates $\ket{ii}$
and Schmidt-rank-two eigenstates $\ket{\psi^\pm_{ij}}$,
respectively. Thus, either $\psi$ is factorized, i.e. there is
only one non-vanishing Schmidt coefficient (equal to 1), or all
the eigenvalues of $(\pro{\psi})^\Gamma$ have modulus strictly
less than one. As regards realignment, we have
$R(\pro{\psi})=\sum_{ij}\mu_i\mu_j\pro{i}\otimes\pro{j}$. For any
pure state $\psi$, thus,
$\|(\pro{\psi})^{\Gamma_A}\|_1=\|R(\pro{\psi})\|_1=(\sum_i\mu_i)^2$.
Therefore, both partial transposition and realignment detect all
pure entangled (bipartite) states.

\subsection{Entanglement witnesses}

It is  well known that any bipartite entangled state $\rho_{AB}$
can be detected by means of a suitable entanglement
witness~\cite{sep1996,TerhalBell2000}: for every entangled state
$\rho_{AB}$ there exists an observable $W=W_{AB}$ such that
$\Tr(W\rho_{AB})<0$, while  $\Tr(W\sigma_{\rm sep})\geq0$ for all
separable states $\sigma_{\rm sep}$. It is clear that a nontrivial
entanglement witness, i.e. an observable able to detect at least
some entangled state, is not positive semidefinite.

A witness  is decomposable~\cite{optwitness} if it can be written
as $W=P+Q^\Gamma$, with $P,Q\geq0$ positive semidefinite
operators. To detect PPT entangled states a witness must be
nondecomposable. Indeed, $\Tr(W\rho)\geq0$ for all PPT state
$\rho$ and all decomposable witnesses $W$.

In~\cite{Schmidtnumber,Schmidtwitness} the concept of
Schmidt-number witness was introduced. A (nontrivial)
Schmidt-number $k$ witness $W$ is an observable which is positive
semidefinite with respect to (mixed) states of Schmidt number
$k-1$, but such that there exists a Schmidt-number $k$ state
$\rho$ such that $\Tr(W\rho)<0$. Notice that standard entanglement
witnesses correspond to Schmidt-number 2 witnesses.

Moreover,  witnesses are able to distinguish between different
kinds of multipartite entanglement~\cite{acin-mixed-3}. Indeed,
there always exists an observable whose expectation value is able
to discriminate between states in a convex subset and a state
outside it. Therefore, for any state $\rho$ that is
$(k+1)$-partite entangled there exists a witness $W$ such that
$\Tr(W\rho)<0$ while $\Tr(W\sigma)\geq0$ for all states $\sigma$
that are at most $k$-partite entangled. Similarly, there is always
a witness which distinguishes a state that is not $k$-separable
from states that are so. In particular, for an $N$-partite state
that is $N$-partite entangled, there exists a witness which tells
it from biseparable states.

\subsection{Maps and entanglement}

A linear map $\Lambda:M_d\rightarrow M_{d'}$ is: \emph{positive}
if $\Lambda[X]\geq0$ for all $X\geq0$; \emph{$k$-positive} if
$\idmap_k\otimes\Lambda$ is positive, with $\idmap_k$ the identity
map on $M_k$; \emph{completely positive} if it is $k$-positive for
all $k\geq1$. It is remarkable that $\Lambda:M_d\rightarrow
M_{d'}$ is completely positive if and only if it is
$d$-positive~\cite{choi}.

Operators $W$  in $M_{dd'}\cong M_d\otimes M_{d'}$ are isomorphic
to linear maps $\Lambda:M_d\rightarrow M_{d'}$, through the
Choi-Jamio{\l}kowski isomorphism~\cite{choi,Jamiolkowski}:
\begin{align}
W&=W_\Lambda=d(\idmap_d\otimes\Lambda)[P_d^+]\\
\label{eq:isomap}\Lambda[X]&=\Lambda_W[X]=\Tr_1\big((X^T\otimes\idmat)W\big),
\end{align}
where the trace in \eqref{eq:isomap} is on the the first subsystem only, and
\beq
\label{eq:maxentpure}
P^+_d\equiv P_{\Psi^+_d},\quad\ket{\Psi^+_d}=\frac{1}{\sqrt{d}}\sum_i\ket{i\otimes i}
\eeq
is the maximally entangled state for a $d\otimes d$ system, $d\geq2$.

In particular,  (nontrivial) witnesses are isomorphic to PnCP
maps. An example of PnCP map is transposition, that fails already
to be $2$-positive, and is associated to $V=d(\idmap_d\otimes
T)[P_d^+]$, that is the swap operator:
$V\ket{\phi\otimes\chi}=\ket{\chi\otimes\phi}$. In the same way as
there is always an entanglement witness which detects a bipartite
entangled states $\rho_{AB}$, there is also a PnCP map $\Lambda$
such  that \beq \label{eq:mapcond}
(\idmap_A\otimes\Lambda_B)[\rho_{AB}]\geq0 \eeq is \emph{not}
satisfied~\cite{sep1996}.

Every nondecomposable  witness is associated to a non-decomposable
map~\cite{optwitness}. A map $\Lambda$ is decomposable if it can
be written as $\Lambda=\Lambda^{\rm CP}_1+\Lambda^{\rm CP}_2\circ
T$, where $\Lambda^{\rm CP}_{1(2)}$ is a completely positive map
and $\circ$ stands for composition. Indeed, Eq. \eqref{eq:mapcond}
is satisfied for all PPT states and decomposable maps. Moreover,
every (non-trivial) Schmidt-number $k$ witness is associated to a
$(k-1)$-positive but not $k$-positive
map~\cite{Schmidtnumber,Schmidtwitness}.

\section{The basic set of states}

We start by considering  a bipartite system with associated
Hilbert space
$\mathcal{H}_{AB}=\mathcal{H}_A\otimes\mathcal{H}_B$, with \beq
\mathcal{H}_A=\mathcal{H}_{A_1}\otimes\mathcal{H}_{A_2},\quad
\mathcal{H}_B=\mathcal{H}_{B_1}\otimes\mathcal{H}_{B_2} \eeq and
$\mathcal{H}_{A_i}=\mathcal{H}_{B_i}=\mathbb{C}^{d_i}$.

We focus on states \beq
\rho_{AB}(\psi^{(1)},\psi^{(2)})=\rho_{1}(\psi^{(1)})\otimes\rho_{2}(\psi^{(2)}),
\eeq with \beq
\rho_i(\psi^{(i)})=\rho_{A_iB_i}=\mathcal{N}_i(\idmat-P_{\psi^{(i)}})_{A_iB_i},
\eeq where $\mathcal{N}_i=1/(d_i^2-1)$ are normalization factors.
Each pure state $\ket{\psi^{(i)}}\equiv\ket{\psi^{(i)}}_{A_iB_i}$
is given by \beq \label{eq:schmidt} \ket{\psi^{(i)}}
=\sum_{j=1}^{r_i}\mu^{(i)}_j\ket{j_{A_i}\otimes j_{B_i}}, \eeq
where \eqref{eq:schmidt}  and $r_i$ are the corresponding Schmidt
decomposition and Schmidt number, respectively. The states
$\rho_i(\psi^{(i)})$ are $A_i:B_i$
separable~\cite{maximally-disordered}, thus
$\rho_{AB}(\psi^{(1)},\psi^{(2)})$ is $A:B$ separable. The partial
transposition of $\rho_i(\psi^{(i)})$ with respect to $A_i$ is:
\beq
\rho_i^{\Gamma_A}(\psi^{(i)})=\mathcal{N}_i(\idmat-(\pro{\psi^{(i)}})^{\Gamma_A}).
\eeq From what we have seen about partial transposition of pure
states in Section \ref{sec:PTrealigment}, it is clear that
$\rho_i^{\Gamma_A}(\psi^{(i)})$ has full rank if and only if
$\psi^{(i)}$ is entangled. It follows that, if both $\psi^{(1)}$
and $\psi^{(2)}$ are entangled,
$\rho_{AB}^{\Gamma_A}(\psi^{(1)},\psi^{(2)})$ is strictly
positive. This implies that a change of
$\rho_{AB}(\psi^{(1)},\psi^{(2)})$ small enough cannot spoil the
positivity of the partial transpose.

This leads us to define the class of states of interest consisting
of the convex combination \eqref{eq:structstate}.  Such class has already
appeared in literature~\cite{boundishizaka,IPppt}, and it was
proved that some states in the class are entangled even if PPT, by
showing that they allow to perform tasks that are impossible under
LOCC. Here, in a different vein, we will look for entanglement
witnesses to prove that, for almost all choices of entangled
states $\psi^{(1)}$ and $\psi^{(2)}$,  the state
$\rho_p(\psi^{(1)},\psi^{(2)})$ is entangled as soon as $p>0$.
Therefore, if $p$ is chosen to be small enough,
$\rho_p(\psi^{(1)},\psi^{(2)})$ is a PPT entangled state. We
remark that if one of  the two states $\psi^{(i)}$ is separable,
while the other is entangled, then $\rho_p(\psi^{(1)},\psi^{(2)})$
is always NPT; $\rho_p(\psi^{(1)},\psi^{(2)})$ can be made PPT
entangled, for some choice of $p>0$, only if \emph{both} pure
states $\psi^{(i)}$ are entangled.

\section{Canonical witness}

\begin{figure}
\includegraphics[width=0.3\textwidth]{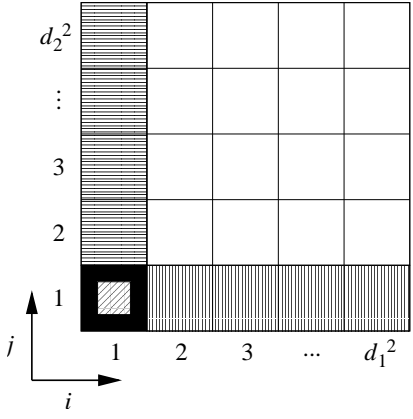}
\caption{Graphical representation of the choice of $p_{ij}$ (white
and black) and $w_{ij}$ (patterns) in \eqref{eq:structstate} and
\eqref{eq:structwitness}, respectively. White color corresponds to
the separable part of $\rho_p(\psi^{(1)},\psi^{(2)})$, while the
vertical and horizontal patterns correspond to the positive part
of the witness $W_\epsilon(\psi^{(1)},\psi^{(2)})$. Black color
and the diagonal pattern stand for $p_{11}=p>0$ and
$w_{11}=-\epsilon<0$, respectively (see the main text for details).} \label{fig:statewitness} 
\end{figure}

We now construct a suitable entanglement witness. If we complete
$\psi^{(i)}$ to a basis
$\{\psi^{(i)}_1\equiv\psi^{(i)},\psi^{(i)}_2,\ldots,\psi^{(i)}_{d_i^2}\}$
of $\mathbb{C}^{d^2_i}$, for both $i=1,2$, we see that
$\rho_p(\psi^{(1)},\psi^{(2)})$ is diagonal in the basis
$\{\psi^{(1)}_i\otimes\psi^{(2)}_j\}$: \beq \label{eq:statediag}
\rho_p(\psi^{(1)},\psi^{(2)})=\sum_{i=1}^{d_1^2}\sum_{j=1}^{d_2^2}p_{ij}P_{\psi^{(1)}_i}\otimes
P_{\psi^{(2)}_j} \eeq with $p_{ij}\geq0$. Note that, since the
tensor product structure is along the $A_1B_1:A_2B_2$ cut, and not
along the $A_1A_2:B_1B_2$ cut, the expression \eqref{eq:statediag}
is not related to the $A:B$ separability property given by
\eqref{eq:sepstates}. We will consider witnesses diagonal in the
same basis, i.e. \beq \label{eq:witnessdiag}
W=\sum_{i=1}^{d_1^2}\sum_{j=1}^{d_2^2}w_{ij}P_{\psi^{(1)}_i}\otimes
P_{\psi^{(2)}_j}, \eeq so that
$\Tr\big(W\rho_p(\psi^{(1)},\psi^{(2)})\big)=\sum_{ij}p_{ij}w_{ij}$.
Of course, the operator \eqref{eq:witnessdiag} is not a trivial
witness only if $w_{ij}<0$ for some $(i,j)$. In particular, if we
define \beq
\begin{split}
\label{eq:structwitness}
W_\epsilon(\psi^{(1)},\psi^{(2)})&=P_{\psi^{(1)}}\otimes(\idmat-P_{\psi^{(2)}})\\
&+(\idmat-P_{\psi^{(1)}})\otimes P_{\psi^{(2)}}\\
&-\epsilon P_{\psi^{(1)}}\otimes P_{\psi^{(2)}}\\
&=P_{\psi^{(1)}}\otimes\idmat+\idmat\otimes P_{\psi^{(2)}}\\
&-(2+\epsilon)P_{\psi^{(1)}}\otimes P_{\psi^{(2)}},
\end{split}
\eeq with $\epsilon\geq0$, we have \beq
\Tr(W_\epsilon(\psi^{(1)},\psi^{(2)})\rho_p(\psi^{(1)},\psi^{(2)}))=-p\epsilon.
\eeq Indeed, $\rho_p(\psi^{(1)},\psi^{(2)})$ and
$W_\epsilon(\psi^{(1)},\psi^{(2)})$ are diagonal in the same basis
and their supports are orthogonal except for the unidimensional
subspace spanned by $\psi^{(1)}\otimes\psi^{(2)}$. A graphical
representation of both the state and the witness decompositions
(in terms of $p_{ij}$ and $w_{ij}$) for the choices
\eqref{eq:structstate} and \eqref{eq:structwitness} is given in
Fig. \ref{fig:statewitness}.

We have to prove that, at least for some choices of $\psi^{(i)}$,
there exists $\epsilon>0$ such that
$W_\epsilon(\psi^{(1)},\psi^{(2)})$ is a non-trivial entanglement
witness.  Indeed, as soon as $\epsilon>0$,
$W_\epsilon(\psi^{(1)},\psi^{(2)})$ is not a positive semidefinite
operator. We proceed by finding the conditions for which it is
positive on separable states:
$\bra{\alpha_A\otimes\beta_B}W_\epsilon(\psi^{(1)},\psi^{(2)})\ket{\alpha_A\otimes\beta_B}\geq
0$, for all factorized (not necessarily normalized)
$\ket{\alpha_A\otimes\beta_B}$. Let us therefore consider vectors
\begin{align*}
\ket{\alpha}&=\sum_{i=1}^{d_1}\sum_{j=1}^{d_2}\alpha_{ij}\ket{i_{A_1}\otimes j_{A_2}}\\
\ket{\beta}&=\sum_{i=1}^{d_1}\sum_{j=1}^{d_2}\beta_{ij}\ket{i_{B_1}\otimes j_{B_2}},
\end{align*}
where $\alpha=[\alpha_{ij}]$, $\beta=[\beta_{ij}]$ are complex
$d_1\times d_2$ rectangular matrices, and where we have taken the
bases $\{\ket{i_{A_k}}\}$, $\{\ket{i_{B_k}}\}$ in the Hilbert
spaces $\mathcal{H}_{A_k}$, $\mathcal{H}_{B_k}$, $k=1,2$ to be the
ones corresponding to the Schmidt decomposition \eqref{eq:schmidt}
of $\psi^{(1)}$ and $\psi^{(2)}$. We find
\begin{multline}
\label{eq:mainrelation}
\bra{\alpha_A\otimes\beta_B}W_\epsilon(\psi^{(1)},\psi^{(2)})\ket{\alpha_A\otimes\beta_B}=\\
\Tr\Big((\beta^T \mu^{(1)}\alpha)^\dagger(\beta^T \mu^{(1)}\alpha)\Big)
+\Tr\Big((\alpha \mu^{(2)}\beta^T)^\dagger (\alpha \mu^{(2)}\beta^T)\Big)\\
-(2+\epsilon)\Big|\Tr(\alpha \mu^{(2)}\beta^T \mu^{(1)}\Big|^2,
\end{multline}
with $\mu^{(i)}=(\mu^{(i)})^\dagger=(\mu^{(i)})^T$ the positive
diagonal matrix of the Schmidt coefficients of $\psi^{(i)}$.

Let us consider a matrix orthonormal basis (o.n.b.) in $M_d$, i.e.
a set of matrices  $\{F_i\}_{i=1}^{d^2}$ such that the matrices
are orthonormal with respect to the Hilbert-Schmidt inner product:
$\Tr(F_i^\dagger F_j)=\delta_{ij}$. For any matrix o.n.b.
$\{F_i\}$ and any matrix $X$, we have $X=\sum_i\Tr(F_i^\dagger
X)F_i$, and $\sum_i|\Tr(F_i^\dagger X)|^2=\Tr(X^\dagger X)$. As
$\Tr({\mu^{(i)}}^2)=1$, each $\mu^{(i)}$ can be be considered as
an element of a matrix o.n.b  in $M_{d_i}$. Let us complete each
$G^{(i)}_1=\mu^{(i)}$ to an o.n.b. $\{G^{(i)}_j\}_{j=1}^{d_i^2}$.
Then
\begin{subequations}
\label{eq:inequalityonb}
\beq
\label{eq:inequalityonb1}
\begin{aligned}
\Big|\Tr(\alpha \mu^{(2)}\beta^T \mu^{(1)})\Big|^2&\leq
\sum_{j=1}^{d_1^2}\Big|\Tr(\alpha \mu^{(2)}\beta^T G^{(1)}_j) \Big|^2\\
&=\Tr\Big((\alpha \mu^{(2)}\beta^T)^\dagger (\alpha \mu^{(2)}\beta^T)\Big),
\end{aligned}
\eeq
and, similarly,
\beq
\label{eq:inequalityonb2}
\begin{aligned}
\Big|\Tr(\beta^T \mu^{(1)}\alpha\mu^{(2)})\Big|^2&\leq
\sum_{j=1}^{d_2^2}\Big|\Tr(\beta^T \mu^{(1)}\alpha G^{(2)}_j) \Big|^2\\
&=\Tr\Big((\beta^T \mu^{(1)}\alpha)^\dagger(\beta^T \mu^{(1)}\alpha)\Big).
\end{aligned}
\eeq
\end{subequations}
Inequalities  \eqref{eq:inequalityonb} correspond to
$P_{\psi^{(1)}}\otimes P_{\psi^{(2)}}\leq \idmat\otimes
P_{\psi^{(2)}}$ and $P_{\psi^{(1)}}\otimes P_{\psi^{(2)}}\leq
P_{\psi^{(1)}}\otimes\idmat$, respectively. Yet having cast them
in the form \eqref{eq:inequalityonb} allows us to argue about the
necessary and sufficient conditions on $\psi^{(1)}$ and
$\psi^{(2)}$ to have a non-trivial witness
$W_\epsilon(\psi^{(1)},\psi^{(2)})$, i.e. to have $\epsilon>0$.

Positivity on factorized states imposes $\epsilon=0$ if and only
if there are matrices $\alpha$ and $\beta$ such that the
inequalities \eqref{eq:inequalityonb} are both saturated at the
same time, i.e. both sums in \eqref{eq:inequalityonb} reduce to
just the first term, and this term does not vanish. Indeed, under
these conditions, \eqref{eq:mainrelation} is equal to
$-\epsilon\Big|\Tr(\alpha \mu^{(2)}\beta^T \mu^{(1)}\Big|^2$ and
strictly negative as soon as $\epsilon>0$. The two sums reduce to
the first term if and only if there are $\alpha$ and $\beta$ such
that~\cite{wlog} \beq \label{eq:system0}
\alpha\mu^{(2)}\beta^T=\mu^{(1)}=G^{(1)}_1,
\qquad\beta^T\mu^{(1)}\alpha=c\mu^{(2)}=cG^{(2)}_1, \eeq where $c$
is a complex constant of proportionality. Only in this case, in
fact, $\alpha\mu^{(2)}\beta^T$ and $\beta^T\mu^{(1)}\alpha$ are
orthogonal to all the other elements of the two matrix o.n.b.. If
condition \eqref{eq:system0} is satisfied, the first two terms on
the right side of \eqref{eq:mainrelation} must be equal. Thus one
finds $|c|=1$ and finally, taking into account Hermiticity and
positivity of $\mu^{(i)}$, one obtains $c=1$.

We have reduced the problem  of determining the existence of a
non-trivial witness of the form \eqref{eq:structwitness}, to that
of verifying whether, for given states $\psi^{(1)}$ and
$\psi^{(2)}$, there exist matrices $\alpha$ and $\beta$ which
solve the system of matrix equations
\begin{subequations}
\label{eq:system}
\begin{align}
\alpha\mu^{(2)}\beta^T&=\mu^{(1)}\\
\qquad\beta^T\mu^{(1)}\alpha&=\mu^{(2)}.
\end{align}
\end{subequations}
First, we notice that this  is possible only if $\mu^{(2)}$ and
$\mu^{(1)}$ have the same rank, i.e. only if the states
$\psi^{(1)}$ and $\psi^{(2)}$ have the same Schmidt number
$r=r_1=r_2$. It is sufficient to focus on this case. We further
observe that w.l.o.g. we can consider the $r$ non-vanishing
Schmidt coefficients of $\psi^{(i)}$ to appear in the first $r$
diagonal entries of $\mu^{(i)}$, for $i=1,2$. Therefore, we can
consider all the matrices entering \eqref{eq:system} to be
$r\times r$ square matrices, even if the initial dimensions $d_1$
and $d_2$ were different. Moreover, they are all invertible, since
we are considering the case the rank of both Schmidt coefficient
matrices $\mu^{(i)}$ is $r$. We can therefore rewrite
\eqref{eq:system} as $\mu^{(2)}\beta^T=\alpha^{-1}\mu^{(1)}$ and
$\beta^T\mu^{(1)}=\mu^{(2)}\alpha^{-1}$. Taking into account that
both matrices $\mu^{(i)}$ are diagonal and strictly positive (i.e.
all the $r$ Schmidt coefficients are not null), we arrive at the
following relations:
\begin{subequations}
\begin{align}
\label{eq:2system1}(\alpha^{-1})_{ij}&=\frac{\mu^{(2)}_i}{\mu^{(1)}_j}\beta_{ji},\\
\label{eq:2system2}\beta_{ji}(\mu^{(1)}_j)^2&=\beta_{ji}(\mu^{(2)}_i)^2.
\end{align}
\end{subequations}
From \eqref{eq:2system1}, we  have that if $\beta_{ji}=0$ then
also $(\alpha^{-1})_{ij}=0$; from $\eqref{eq:2system2}$ we find
that, if $\beta_{ji}\neq 0$, then $\mu^{(1)}_j=\mu^{(2)}_i$ and
therefore, from $\eqref{eq:2system1}$,
$(\alpha^{-1})_{ij}=\beta_{ji}$. In conclusion, we have
$\alpha^{-1}=\beta^T$. Therefore, a solution to equations
\eqref{eq:system} exists only if $\mu^{(1)}$ and $\mu^{(2)}$ are
connected by a similarity transformation \beq
\label{eq:similarity} \alpha \mu^{(2)} \alpha^{-1}=\mu^{(1)}, \eeq
and have the same eigenvalues. In such case, we have that
\eqref{eq:mainrelation} reduces to $-\epsilon
\Tr((\mu^{(i)})^2)=-\epsilon$, so that we must choose $\epsilon=0$
to have positivity on separable states.

We have shown that the  witness
$W_\epsilon(\psi^{(1)},\psi^{(2)})$ defined in
\eqref{eq:structwitness} can always be chosen to be non-trivial,
i.e. with $\epsilon>0$, except in the case $\psi^{(1)}$ and
$\psi^{(2)}$ have essentially the same Schmidt decomposition.
Notice that w.l.o.g. we can consider the Schmidt coefficients to
be ordered as $\mu^{(i)}_k\geq\mu^{(i)}_{k+1}$, for $i=1,2$. Thus,
we have always a witness except in the case $\mu^{(1)}=\mu^{(2)}$
(indeed, the similarity transformation \eqref{eq:similarity} is
actually a permutation). Correspondingly, we have proved that, for
almost all pairs of pure entangled states $\psi^{(i)}$, $i=1,2$,
the state $\rho_p(\psi^{(1)},\psi^{(2)})$ is entangled as soon as
$p>0$.

\section{Partial transposition and realignment}
\label{sec:PTandReal}

Let us now consider more  in detail the behavior of the class of
states $\rho_p(\psi^{(1)},\psi^{(2)})$ under the operations of
partial transposition and realignment~\cite{footnote1}. Partial
transposition for such states has already been studied
in~\cite{boundishizaka}. For completeness, we reproduce here those
results, and we extend the analysis by comparing the entanglement
detection power of partial transposition and realignment.
Moreover, we observe that no element in the class is a candidate
to be an NPT bound state, i.e. as soon as the states are NPT, they
are provably distillable.

\begin{figure}[h]
\begin{center}
\includegraphics[angle=0,width=0.4\textwidth]{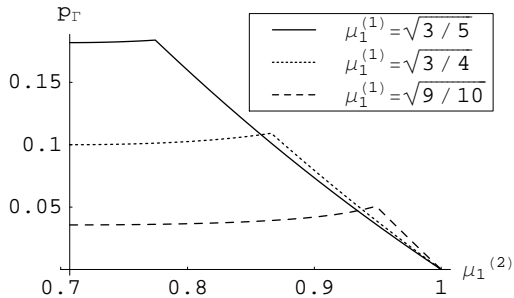}
\caption{Dependence of the threshold probability $p_\Gamma$ on
$\psi^{(2)}$ for fixed choices of $\psi^{(1)}$ in the case
$d_1=d_2=2$. The state $\rho_p(\psi^{(1)},\psi^{(2)})$ is entangled 
for all choices of $\mu_1^{(1)}\neq\mu_1^{(2)}$; when $0\leq p<p_\Gamma$ the state is
PPT, while it is NPT if  $p_\Gamma< p\leq1$. 
The point in which the minimum on the right hand side of
(\ref{eq:PPTcondition}) changes from one element to the other is
clear from the sharp change in the behavior of the curve, and it 
coincides with the point $\mu_1^{(1)}=\mu_1^{(2)}$ (note that, for such point, with the 
methods introduced in this work we are
not able to say that the state $\rho_p(\psi^{(1)},\psi^{(2)})$ is entangled when PPT 
(see main text and Table \ref{tab:entprop}).
As expectable, the threshold value of $p_\Gamma$ goes to 0
as one of the two states becomes less entangled.}\label{fig:d1d22}
\end{center}
\end{figure}

As regards PT, we have  that the eigenvalues of
$\rho^{\Gamma_A}_p(\psi^{(1)},\psi^{(2)})$ are
$(1-p)\mathcal{N}_1\mathcal{N}_2(1-\lambda^{(1)})(1-\lambda^{(2)})+p\lambda^{(1)}\lambda^{(2)}$,
where the $\lambda^{(i)}$s run over eigenvalues of
$(\pro{\psi^{(i)}})^\Gamma$, $i=1,2$. Let us recall that a state
$\rho_{AB}$ is distillable if and only if there exist a number of
copies $n$ and a Schmidt rank 2 state $\phi_2$ such that
$\bra{\phi_2}(\rho^{\Gamma_A})^{\otimes
n}\ket{\phi_2}<0$~\cite{bound}. It is easy to see that the minimum
eigenvalue of $\rho^{\Gamma_A}_p(\psi^{(1)},\psi^{(2)})$ is of the
form
\beq
\label{eq:smallesteigenvalue}
(1-p)\mathcal{N}_1\mathcal{N}_2(1-(\mu^{(i)}_k)^2)
(1+\mu^{(j)}_m\mu^{(j)}_n)-p(\mu^{(i)}_k)^2\mu^{(j)}_m\mu^{(j)}_n,
\eeq
with $m\neq n$ and $(i,j)\in\{(1,2),(2,1)\}$, i.e. it
corresponds to a Schmidt rank 2 eigenvector
$\ket{k_{A_i}k_{B_i}}\otimes\ket{\psi_{mn}^-}_{A_jB_j}$.
Therefore, as soon as the state is NPT, we prove that it is also
distillable by considering $n=1$ and taking as $\phi_2$ the
eigenvector corresponding to the minimal negative eigenvalue. On
the other hand, by choosing $p$ small enough, it always possible
to make the smallest eigenvalue positive, if the first term in
\eqref{eq:smallesteigenvalue} is not null, i.e. if \emph{both}
states $\psi^{(i)}$ are entangled.  More precisely, it can be
shown~\cite{boundishizaka} that the necessary and sufficient
condition for the state to be PPT is \beq
\begin{aligned}
\label{eq:PPTcondition}
\frac{p}{(1-p){\mathcal{N}}_1{\mathcal{N}}_2}\leq \min \Big\{
&\frac{(1-(\mu_1^{(1)})^2)(1+\mu_1^{(2)}\mu_2^{(2)})}{(\mu_1^{(1)})^2\mu_1^{(2)}\mu_2^{(2)}},\\
&\frac{(1-(\mu_1^{(2)})^2)(1+\mu_1^{(1)}\mu_2^{(1)})}{(\mu_1^{(2)})^2\mu_1^{(1)}\mu_2^{(1)}}
\Big\}.
\end{aligned}
\eeq
In particular, to calculate the smallest eigenvalue of the
partially-transposed state, it is sufficient to consider only the
two biggest Schmidt coefficients of $\psi^{(1)}$ and $\psi^{(2)}$. 
We will denote by $p_\Gamma$ the largest value of $p$ for which
$\rho_p(\psi^{(1)},\psi^{(2)})$ is PPT. In Fig. \ref{fig:d1d22}  we 
plot the dependence of $p_\Gamma$ on the Schmidt coefficients of 
the two pure states in the case $d_1=d_2=2$ (i.e. when $\rho_p(\psi^{(1)},\psi{(2)})$ is a state of four qubits).

The condition to  determine when the realignment criterion detects
entanglement is not trivial to handle analytically. Thus, we will restrict ourselves to
the case in which the two pure states $\psi^{(1)}$ and
$\psi^{(2)}$ are maximally entangled. In this case, we have
\beq
\begin{split}
R(\rho_p(\Psi^+_{d_1},\Psi^+_{d_2}))&=(1-p)\mathcal{N}_1
\mathcal{N}_2\left(\sum_{i=1}^{d_1}\ket{ii}\bra{jj}-\frac{\idmat}{d_1}\right)\otimes\\
&\left(\sum_{i,j=1}^{d_2}\ket{ii}\bra{jj}-\frac{\idmat}{d_2}\right)+p\frac{\idmat}{d_1d_2}.
\end{split}
\eeq
The condition $\Vert
R(\rho_p(\Psi^+_{d_1},\Psi^+_{d_2}))\Vert>1$ is thus satisfied
only for  $p>\frac{d_1d_2-2}{d_1^2(d_2^2-2)}$, where we have
assumed w.l.o.g. $d_2\geq d_1$. Note that this value is always
greater than $p_\Gamma=\frac{1}{1+(d_1+1)(d_2-1)}$: thus, in the
case in which the pure states $\psi^{(1)}$ and $\psi^{(2)}$ are
maximally entangled, realignment is always less sensitive than PT.

\begin{figure}[t]
\begin{center}
\includegraphics[angle=0,width=0.4\textwidth]{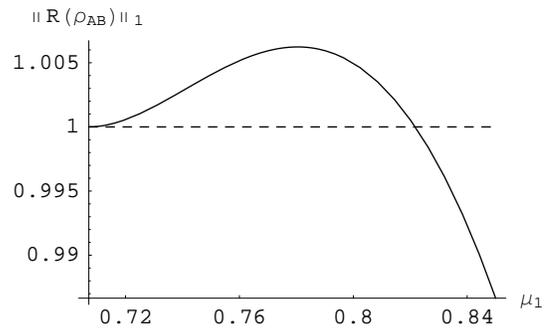}
\caption{Comparison of the detection power of realignment and partial transposition in the $d_1=d_2=2$ case. We take $\psi^{(1)}=\psi^{(2)}=\psi$, where $\psi$ is a pure (entangled) state of two qubits, characterized by its larger Schmidt coefficient $\mu_1$. We consider $\|R(\rho_{AB})\|_1$, for $\rho_{AB}=\rho_{p_\Gamma}(\psi,\psi)$, i.e. for the state at the border of PPT states. For most of the range $1/\sqrt{2}\leq\mu_1\leq 1$ partial transposition is more sensitive than realignment, i.e. $\|R(\rho_{AB})\|_1<1$ even if a slight change of $p$ makes the state NPT entangled. Anyway the plot shows that realignment is more sensitive than partial transposition for $\psi$ almost maximally entangled, i.e $\|R(\rho_{AB})\|_1>1$ even if the state is PPT.} 
\label{fig:graphRealPgamma}
\end{center}
\end{figure}

In Section \ref{sec:morethantwo} we will provide analytical examples of
states, which have a structure similar (see Eq. \eqref{eq:genstates}) to that of $\rho_p(\psi^{(1)},\psi^{(2)})$, detected as entangled by realignment but not by partial transposition. In Fig. \ref{fig:graphRealPgamma} we show that realignment and partial transposition are inequivalent (i.e., there are entangled states detected by one criterion but not by the other one) also in the class $\rho_p(\psi^{(1)},\psi^{(2)})$. The plot of Fig. \ref{fig:graphRealPgamma} is relevant also for another reason: it shows that also states for which it is not possible to construct a non-trivial witness $W_\epsilon(\psi^{(1)},\psi^{(2)})$ (i.e., states for which $\mu^{(1)}=\mu^{(2)}$) may be entangled.

For the sake of clarity, in Table \ref{tab:entprop} we summarize the relation between the entanglement properties
of the two pure states $\psi^{(1)},\psi^{(2)}$, and those of $\rho_p(\psi^{(1)},\psi^{(2)})$.

\begin{table}
\begin{tabular} {|c|c|}
\hline
\hline
$\psi^{(1)}$, $\psi^{(2)}$ &  $\rho_p(\psi^{(1)},\psi^{(2)})$\\
\hline
\hline
both separable
&
separable for all $0\leq p\leq 1$
\\
\hline
one entangled
&
NPT entangled for all $0\leq p\leq 1$
\\
\hline
\multirow{2}{*}{both entangled}
&
$\mu^{(1)}\neq\mu^{(2)}$: PPT entangled for $0<p<p_\Gamma$\\
&
$\mu^{(1)}=\mu^{(2)}$: no general statement
\\
\hline
\end{tabular}
\caption{Relation between the entanglement properties of the two pure states $\psi^{(1)},\psi^{(2)}$, and those of $\rho_p(\psi^{(1)},\psi^{(2)})$. When both $\psi^{(1)},\psi^{(2)}$ are entangled, and do not have the same Schmidt coefficients (i.e., they are not equivalent up to local unitaries), $\rho_p(\psi^{(1)},\psi^{(2)})$ is PPT entangled in the interval $0<p<p_\Gamma$. If both the pure states are entangled, but $\mu^{(1)}=\mu^{(2)}$, the techniques (witnesses) adopted in this work do not help. There are choices of $\psi^{(1)},\psi^{(2)}$ such that the mixed state $\rho_p(\psi^{(1)},\psi^{(2)})$ is separable as soon as, decreasing $p$, it is PPT (see Section \ref{sec:morethantwo}), as well as other choices such that the corresponding mixed states can be PPT entangled (see Fig. \ref{fig:graphRealPgamma}).}
\label{tab:entprop}
\end{table}

\section{Generalization to more than two states $\psi^{(i)}$}
\label{sec:morethantwo}

It is  possible to straightforwardly generalize the construction
of the states  $\rho_p(\psi^{(1)},\psi^{(2)})$ to the case in
which one considers more than two pure states $\psi^{(i)}$.

Given a set of states $\{\psi^{(i)}\}_{i=1}^M$ and a probability
$p$, we define: \beq \label{eq:genstates}
\rho_p(\{\psi^{(i)}\})=(1-p)\bigotimes_{i=1}^M
\frac{\idmat-P_{\psi^{(i)}}}{d_i^2-1}+p\bigotimes_{i=1}^MP_{\psi^{(i)}}.
\eeq To prove that  for $M\geq3$ the state is entangled for $p>0$
as soon as one of the $\psi^{(i)}$ is entangled, it is sufficient
to use the class of witnesses we studied for $M=2$.

\noindent Indeed, for $M\geq3$ it is always possible to split any
set of natural numbers $\{r^{(i)}\}_{i=1}^M$ into two non-empty
disjoint sets, which w.l.o.g can be indicated as
$\{r^{(i)}\}_{i=1}^m$ and $\{r^{(i)}\}_{i=m+1}^M$, and such that
$\prod_{i=1}^mr^{(i)}\neq\prod_{i=m+1}^Mr^{(i)}$. Let us consider
the case in which the numbers $\{r^{(i)}\}_{i=1}^M$ are the
Schmidt ranks of the states in $\{\psi^{(i)}\}_{i=1}^M$. For the
sake of testing entanglement, it is possible to consider two
states $\ket{\tilde{\psi}^{(1)}}=\bigotimes_{i=1}^m\psi^{(i)}$ and
$\ket{\tilde{\psi}^{(2)}}=\bigotimes_{i=m+1}^M\psi^{(i)}$ of
different Schmidt rank (which is a multiplicative quantity under
tensoring). Thus, if at least one state $\psi^{(i)}$ is entangled,
we can construct a non-trivial entanglement witness
$W_\epsilon(\tilde{\psi}^{(1)},\tilde{\psi}^{(2)})$, $\epsilon>0$,
as in \eqref{eq:structwitness} such that
$\Tr(\rho_p(\{\psi^{(i)}\}W_\epsilon(\tilde{\psi}^{(1)},\tilde{\psi}^{(2)}))=-p\epsilon$.
Notice that, if all states $\psi^{(i)}$ are separable, then also
$\rho_p(\{\psi^{(i)}\})$ is separable, i.e. there is no
entanglement to be detected.

Similarly to the case $M=2$, it is possible to prove that the
smallest eigenvalue of a state \eqref{eq:genstates} corresponds to
a Schmidt rank 2 eigenvector, so that as soon as the state is NPT
we know also it is distillable. Moreover, it is possible to find a
$p>0$, such that the state $\rho_p(\{\psi^{(i)}\})$ is PPT
entangled, if and only if the states $\{\psi^{(i)}\}_{i=1}^M$ are
\emph{all} entangled.

\subsection{Maximally entangled pure states $\psi^{(i)}$}

We now focus on a even more specific class of states. Recalling
the definition  \eqref{eq:maxentpure} of maximally entangled state
$P^+_{d}$, we define the states
\beq
\label{eq:maxent}
\rho_p(d_1,\ldots,d_M)=(1-p)\bigotimes_{i=1}^M\frac{\idmat-P^+_{d_i}}{d_i^2-1}+p\bigotimes_{i=1}^MP^+_{d_i}.
\eeq
Compare them to the isotropic states for a $d\otimes d$
system: \beq \rho_p(d)=(1-p)\frac{(\idmat-P^+_d)}{d^2-1}+pP^+_d.
\eeq Isotropic states can be considered a subclass of the class of
states we are studying, with $M=1$. It is remarkable that
isotropic states $\rho_p(d)$ are either distillable or separable:
no phenomenon of bound entanglement (either PPT or NPT -- if
existing) is present in such class, while it is sufficient to go
to $M=2$ to have it.

It is worth noticing that $\rho_p(d,d)$ is separable for all
values of $p$ for which it is PPT, i.e for $0\leq
p\leq1/d^2$~\cite{entundersimmetry}, and indeed we are not able to
construct a witness of the form \eqref{eq:structwitness} for it,
since in this case $\psi^{(1)}$ and $\psi^{(2)}$ have the same
Schmidt coefficients -- they are equal. On the other hand, a
witness as in \eqref{eq:structwitness} exists for
$\rho_p(d_1,d_2)$ in the case $d_2>d_1\geq2$.

As regards the sensitivity of partial transposition, for $\rho_p(d_1,\ldots,d_M)$ we have
\[
p_\Gamma=\frac{1}{1+(d_M-1)\prod_{i=1}^{M-1}(d_i+1)},
\]
taking w.l.o.g. $d_1\leq d_2\leq\ldots\leq d_M$.
We have seen that,
if $M=2$ and we consider the case in which both pure states
$\psi^{(i)}$ are maximally entangled in dimension $d_i$, PT is
always more sensitive than realignment. To study the more general
case $M>2$, we restrict ourselves, for simplicity, to the case in
which all the dimensions $d_i$ coincide ($d_i=d\geq2$) and
$\psi^{(i)}=\Psi^+_d$ for all $i=1,\ldots M$: \beq
\rho_p(d;M)\equiv\rho_p(\underbrace{d,d,\ldots,d}_{\textrm{M
times}}). \eeq In this case we have that $\rho_p(d;M)$ is PPT for
$p\leq p_\Gamma=\frac{1}{1+(d-1)(d+1)^{M-1}}$. As regards
realignment, we have \beq \label{eq:realignmentvalue} \Vert
R(\rho_p(d;M))\Vert_1=\frac{1}{d^M}\sum_{j=0}^M\binom{M}{j}\vert1-p-p(1-d^2)^j\vert~.
\eeq While it is not trivial to find an analytical solution in $p$
of the inequality $\Vert R(\rho_p(d;M))\Vert_1\leq1$, it is
possible to see that there are cases in which the realignment
criterion is more sensitive than PT. Indeed, this happens for
$d=2$ and $M\geq3$ odd. To verify this, it is sufficient to plug
in \eqref{eq:realignmentvalue} the corresponding value of
$p_\Gamma$, i.e. $p=p_\Gamma=\frac{1}{1+3^{M-1}}$. By definition,
for such value of $p$ the state is PPT and the condition $\Vert
R(\rho_{p_\Gamma}(d;M))\Vert_1>1$ is satisfied for all odd values
of $M\geq3$, while $\Vert R(\rho_{p_\Gamma}(d;M))\Vert_1=1$ for
$M=1$ and $M$ even. Numerical results indicate that $d=2$ and
$M\geq3$ odd is the only case in which realignment detects PPT
entangled states of the form $\rho_p(d;M)$, but we could not
verify this analytically.

\section{More on witnesses}

We proceed now to some remarks as regards the witnesses we analyzed.

\subsection{Simplified witnesses}
\label{sec:simplifiedwitness}

We have seen that the necessary and sufficient condition to have a
non-trivial entanglement witness
$W_\epsilon(\psi^{(1)},\psi^{(2)})$, with $\epsilon>0$, is that
the states $\psi^{(i)}$ have different Schmidt coefficients. When
the Schmidt ranks of states $\psi^{(i)}$ are different, i.e.
w.l.o.g. $r_1<r_2$, it is possible to detect the entanglement of
$\rho_p(\psi^{(1)},\psi^{(2)})$ by means of a witness with a
structure even simpler than that of
$W_\epsilon(\psi^{(1)},\psi^{(2)})$. In such a case, in fact, it
is possible to consider non trivial ($\epsilon>0$) witnesses of
the form \beq
\begin{split}
\label{eq:tildewitness}
\tilde{W}_\epsilon(\psi^{(1)},\psi^{(2)})&=P_{\psi^{(1)}}\otimes(\idmat-(1+\epsilon)P_{\psi^{(2)}})\\
    &=P_{\psi^{(1)}}\otimes\idmat-(1+\epsilon)P_{\psi^{(1)}}\otimes P_{\psi^{(2)}}.
\end{split}
\eeq
For this choice,
\begin{multline}
\label{eq:mainrelation2}
\bra{\alpha_A\otimes\beta_B}\tilde{W}_\epsilon(\psi^{(1)},\psi^{(2)})\ket{\alpha_A\otimes\beta_B}=\\
\Tr\Big((\beta^T \mu^{(1)}\alpha)^\dagger(\beta^T \mu^{(1)}\alpha)\Big)
-(1+\epsilon)\Big|\Tr(\mu^{(2)}\beta^T \mu^{(1)}\alpha)\Big|^2.
\end{multline}
Following the same reasoning we used for
${W}_\epsilon(\psi^{(1)},\psi^{(2)})$, we see that the quantity
\eqref{eq:mainrelation2} can be made negative for any $\epsilon>0$
if and only if (w.l.o.g) there exist $\alpha$ and $\beta$ such
that
\[
\mu^{(2)}=\beta^T \mu^{(1)}\alpha.
\]
This is possible if and only if the rank of $\mu^{(1)}$ is greater than that of $\mu^{(2)}$.

We may better understand this result by considering that
\begin{multline}
\tilde{W}_\epsilon(\psi^{(1)},\psi^{(2)})\\=(P_{\psi^{(1)}}\otimes\idmat)\circ
\Big(\idmat\otimes\big(\idmat-(1+\epsilon)P_{\psi^{(2)}}\big)\Big)\circ(P_{\psi^{(1)}}\otimes\idmat)
\end{multline}
and that
\begin{equation}
\label{eq:basicclarisse}
(P_{\psi^{(1)}}\otimes\idmat)\ket{\alpha_A\otimes\beta_B}=\ket{\psi^{(1)}}\otimes\ket{\gamma},
\eeq with
$\ket{\gamma}=\sum_i\mu_i^{(1)}\Big(\sum_l\alpha_{il}\ket{l}\Big)
\otimes\Big(\sum_k\beta_{ik}\ket{k}\Big)$. It is clear that, by
the right choice of $\alpha$ and $\beta$, $\gamma$ -- though in
general not normalized -- can be made proportional to any state
whose Schmidt rank is not greater than the one of $\psi^{(1)}$. In
particular, if $\psi^{(2)}$ has the same Schmidt rank that
$\psi^{(1)}$ has, it is possible to choose $\alpha$ and $\beta$
such that
\[
\bra{\alpha\beta}\tilde{W}_\epsilon(\psi^{(1)},\psi^{(2)})\ket{\alpha\beta}=-\epsilon |c|^2,
\]
with $\ket{\gamma}=c\ket{\psi^{(2)}}$ and $|c|>0$. Therefore, in
this case, $\tilde{W}$ is positive on separable states if and only
if $\epsilon=0$.

Notice that in Section \ref{sec:morethantwo}, when analyzing the
multi-state case for $M\geq3$, we argued that as soon as one state
$\psi^{(i)}$ is entangled, it is possible to consider two states
$\ket{\tilde{\psi}^{(j)}}$, $j=1,2$ of different rank obtained
from $\psi^{(i)}$s by tensoring. It is therefore clear that for
$M\geq3$, as soon as the problem is not trivial (i.e. not all the
states $\psi^{(i)}$ are factorized), it is always possible to
consider a witness
$\tilde{W}_\epsilon(\tilde{\psi}^{(1)},\tilde{\psi}^{(2)})$ of the
form \eqref{eq:tildewitness}.

\subsection{Canonical witnesses for $\rho_p(\{\psi^{(i)}\})$}
\label{sec:prowitness}

Both for $W=W_\epsilon(\psi^{(1)},\psi^{(2)})$ (Eq.
\eqref{eq:structwitness}) and
$W=\tilde{W}_\epsilon(\psi^{(1)},\psi^{(2)})$ (Eq.
\eqref{eq:tildewitness}), we have not only
$\Tr(W\rho_0(\psi^{(1)},\psi^{(2)}))=0$, but, more strongly,
\[
W\rho_0(\psi^{(1)},\psi^{(2)})=\rho_0(\psi^{(1)},\psi^{(2)})W=0,
\]
i.e. the witnesses \cite{footnote3} are orthogonal to the
separable part -- which corresponds to
$\rho_0(\psi^{(1)},\psi^{(2)})$ -- of a state
$\rho_p(\psi^{(1)},\psi^{(2)})$. Indeed, we have
$\tilde{W}_\epsilon(\psi^{(1)},\psi^{(2)})\leq
W_\epsilon(\psi^{(1)},\psi^{(2)})$ (compare
\eqref{eq:structwitness} and \eqref{eq:tildewitness}) and
\[
\begin{split}
W_\epsilon(\psi^{(1)},\psi^{(2)})&=\idmat-(\idmat-P_{\psi^{(1)}})\otimes(\idmat-P_{\psi^{(2)}})\\
&-\epsilon P_{\psi^{(1)}}\otimes P_{\psi^{(2)}}.
\end{split}
\]
Moreover, $\rho_0(\psi^{(1)},\psi^{(2)})$ is exactly defined as
the state corresponding (via normalization) to the projector
$(\idmat-P_{\psi^{(1)}})\otimes(\idmat-P_{\psi^{(2)}})$.

In the case of $M\geq3$ states $\psi^{(i)}$, we argued (see
Section \ref{sec:morethantwo}, paragraph following Eq.
\eqref{eq:genstates}) that, as soon as one state $\psi^{(i)}$ is
entangled, there exists a non-trivial entanglement witness of the
form $W_\epsilon(\tilde{\psi}^{(1)},\tilde{\psi}^{(2)})$ which
detects the entanglement of $\rho_p(\{\psi^{(i)}\})$. The states
$\tilde{\psi}^{(i)}$, $i=1,2$ were taken to be tensor products of
two disjoint subsets of $\{\psi^{(i)}\}$, so that
$\ket{\tilde{\psi}^{(1)}}\otimes\ket{\tilde{\psi}^{(2)}}=\bigotimes_{i=1}^M
\ket{\psi^{(i)}}$. We can instead consider a witness of the form
\beq \label{eq:structgenwitness}
\begin{split}
W_\epsilon(\{\psi^{(i)}\})&=\idmat-\bigotimes_{i=1}^M(\idmat-P_{\psi^{(i)}})\\
                                                    &-\epsilon \bigotimes_{i=1}^M P_{\psi^{(i)}}.
\end{split}
\eeq
We have $W_\epsilon(\{\psi^{(i)}\})\geq
W_\epsilon(\tilde{\psi}^{(1)},\tilde{\psi}^{(2)})$, but
$W_\epsilon(\{\psi^{(i)}\})$ has the same expectation value
$-\epsilon p$ with respect to the states $\rho_p(\{\psi^{(i)}\})$.
Moreover, it can be considered as a modification of the projector
onto the subspace orthogonal to the support of the separable part
$\rho_0(\{\psi^{(i)}\})$ of the state, with the modification
$-\epsilon \bigotimes_{i=1}^M P_{\psi^{(i)}}$ tailored to
``intercept'' the entangled part of the state. Notice that the
witness $W_\epsilon(\{\psi^{(i)}\})$ depends only on $\epsilon$
and on the set $\{\psi^{(i)}\}$, not on the choice of two subsets
of  $\{\psi^{(i)}\}$, unlike
$W_\epsilon(\tilde{\psi}^{(1)},\tilde{\psi}^{(2)})$.

\section{Multipartite case}

Now we consider the multipartite case, i.e. the states
$\psi^{(i)}$, $i=,1\ldots,M$, are states of $N$ parties. From the
results presented in Section \ref{sec:PTandReal}, we know that the
state $\rho_p(\{\psi^{(i)}\})$ can be made PPT, with respect to a
given bipartite cut $S_1:S_2$, for some strictly positive $p$ only
if all the states $\psi^{(i)}$ are entangled with respect to that
cut. Therefore, for this to happen for any possible bipartite cut,
all the states $\psi^{(i)}$ must be $N$-partite entangled.

As regards witnesses, we are able to provide a non-trivial (i.e.
not positive semidefinite) witness which detects bipartite
$S_1:S_2$ entanglement, if (i) $M=2$ and the states $\psi^{(1)}$
and $\psi^{(1)}$ have different Schmidt coefficients with respect
to the cut, or (ii) $M\geq3$ and at least one state $\psi^{(i)}$
is entangled with respect to the cut. As we discussed in Section
\ref{sec:prowitness}, it is always possible to consider witnesses
$W_\epsilon(\{\psi^{(i)}\})$ of the form
\eqref{eq:structgenwitness}, for every cut. In the construction of
such witnesses, the only parameter dependent from the cut is
$\epsilon$. If, for a given cut, one of the just mentioned
conditions (i) and (ii) is valid, then it is possible to take
$\epsilon>0$ and detect bipartite entanglement by means of the
corresponding witness. Let us consider
\[
\tilde{\epsilon}=\min_{S_1:S_2}\max\{\,\epsilon\,|\,\bra{\alpha_{S_1}
\otimes\beta_{S_2}}W_\epsilon(\{\psi^{(i)}\})\ket{\alpha_{S_1}\otimes\beta_{S_2}}\geq0\}.
\]
If $\tilde{\epsilon}>0$, then
$W_{\tilde{\epsilon}}(\{\psi^{(i)}\})$ is a nontrivial witness for
genuine multipartite entanglement. Thus, we get immediately that
$\rho_p(\{\psi^{(i)}\})$ is $N$-partite entangled for every $p>0$,
since its entanglement is detected by a witness which is positive
with respect to all biseparable states.

Thus, it is possible to decide to construct a state which is PPT
with respect to some desired bipartitions, and NPT with respect to
the remaining ones. To do so, it is sufficient to choose
opportunely the states $\{\psi^{(i)}\}$ and $p$. Indeed, the mixed
state is NPT with respect to a bipartition $S_1:S_2$ for all $p>0$
if and only if there is a state $\psi_i^{(i)}$ which is $S_1:S_2$
separable. If the states satisfy (i) or (ii) for every cut, the
mixed state $\rho_p(\{\psi^{(i)}\})$ for sure contains $N$-partite
entanglement (for $p>0$), because there is a witness which detects
it.

\section{Tensor-like witnesses, $k$-positive maps and nondecomposability}

Building upon the considerations of Section
\ref{sec:simplifiedwitness}, we discuss here the possibility of
obtaining non-decomposable witnesses able to detect PPT bound
entangled states, by composing through tensor product a
decomposable witness (unable to detect PPT bound entanglement)
with a positive operator (w.l.o.g. a state).

Lemma III.1 of \cite{clarisse} says that, if a state $\sigma$ on
$\mathcal{H}_{A_1}\otimes\mathcal{H}_{B_1}$ has Schmidt number
$k$, and $\eta$ is an operator on
$\mathcal{H}_{A_2}\otimes\mathcal{H}_{B_2}$ which is positive with
respect to states of Schmidt number $kl$, then $\sigma\otimes\eta$
is positive with respect to states $\tau$ on
$\mathcal{H}_{A}\otimes\mathcal{H}_{B}$ of Schmidt number $l$,
i.e. $\Tr\big((\sigma\otimes\eta)\tau\big)\geq0$. The proof of
such Lemma \cite{clarisse} is a generalization of the reasoning we
have adopted in Section \ref{sec:simplifiedwitness}. In
particular, it is sufficient to consider the case
$\sigma=\pro{\phi}$ and $\tau=\pro{\psi}$, and notice that
\[
\begin{split}
\Tr\big(\pro{\psi}(\pro{\phi}\otimes\eta)\big)&=\Tr_2\Big(\Tr_1\big(\pro{\psi}(\pro{\phi}\otimes\idmat)\big)\eta\Big)\\
&=\bra{\gamma}\eta\ket{\gamma},
\end{split}
\]
with
$(\pro{\phi}\otimes\idmat)\ket{\psi}=\ket{\phi}\otimes\ket{\gamma}$.
Considering that $\psi$ has Schmidt rank $l$, and that the action
of $(\pro{\phi}\otimes\idmat)$ on a separable states can create at
most a state of Schmidt rank $k$ (see \eqref{eq:basicclarisse}),
we conclude that the state $\gamma$ has at most Schmidt rank $kl$.
Note that $\gamma$ is not normalized, in general.

We are interested in operators that are entanglement witnesses,
i.e. such that they are positive on separable states. We
correspondingly take $l=1$, and consider a state
$\sigma=\pro{\phi}=P_\phi$ of Schmidt number $k$ and a
Schmidt-rank $m$ witness $W$, with $m\geq k+1$. We compose them to
give an operator $P_\phi\otimes W$, which is then positive on
separable states by construction, according to the Lemma III.1 of
\cite{clarisse}. If (i) $\phi$ is entangled (i.e. $k\geq2$) and
(ii) $\psi_W$, of Schmidt rank strictly greater than $k$, is such
that $\bra{\psi_W}W\ket{\psi_W}<0$, then $P_{\phi}\otimes W$ is a
non-decomposable entanglement witness. Indeed, we can consider
$p>0$ such that $\rho_p(\phi,\psi_W)$ is PPT -- both $\phi$ and
$\psi_W$ are entangled -- and have
\[
\Tr\Big(P_\phi\otimes W\rho_p(\phi,\psi_W)\Big)=p\bra{\psi_W}W\ket{\psi_W}<0.
\]

With these considerations and exploiting the Choi-Jamio{\l}kowski
isomorphism, it is immediate to state a theorem relating the
properties of $k$-positivity, complete positivity and
decomposability of maps.
\begin{theorem}
A linear map $\Lambda$ which is $k$-positive, $k\geq2$, is
completely positive if and only if $\idmap_k\otimes\Lambda$ is
decomposable.
\end{theorem}
\begin{proof}
The \emph{only if} part is trivial: if $\Lambda$ is CP, then
$\idmap_k\otimes\Lambda$ is trivially decomposable -- it is CP
itself. To prove the \emph{if} part, let us suppose that $\Lambda$
is not CP and show that $\idmap_k\otimes\Lambda$ is
nondecomposable. Indeed, if $\Lambda$ is not CP, then, even though
the corresponding witness $W_\Lambda$ is positive on Schmidt rank
$k$ states \cite{Schmidtnumber,Schmidtwitness}, there exists a
Schmidt rank $m$ state $\psi_{W_\Lambda}$, $m>k$, such that
$\bra{\psi_{W_\Lambda}}W_\Lambda\ket{\psi_{W_\Lambda}}<0$. Thus,
as remarked before, the witness $P_k^+\otimes W$ is
nondecomposable and the same holds for its isomorphic map
$\idmap_k\otimes\Lambda$.
\end{proof}
Notice that in the theorem we could have used
$\idmap_l\otimes\Lambda$, with any $2\leq l\leq k$, instead of
$\idmap_k\otimes\Lambda$.

The results just exposed imply that, for $k\geq2$, as soon as we
know that a PnCP map $\Lambda$ is $k$-positive or that a
non-positive witness $W$ is positive on Schmidt-number $k$ states,
we know that, for example, $\idmap_k\otimes\Lambda$ and
$P^+_k\otimes W$ are respectively a positive nondecomposable map
and a nondecomposable witness, without caring about the
decomposability of $\Lambda$ or $W$.

We provide now a simple example illustrating how to pass from
provably decomposable witnesses to nondecomposable witnesses
through tensoring. Obviously, the example could be recast
immediately in terms of maps.

Let us consider witnesses of the form
\[
W_\epsilon(\psi)=\idmat-(1+\epsilon)\pro{\psi}.
\]
If we consider expectation values with respect to pure states we obtain
\[
\bra{\varphi} W_\epsilon(\psi) \ket{\varphi}=1-(1+\epsilon)|\langle \psi | \varphi\rangle|^2.
\]
Suppose $\psi$ has Schmidt rank $r$ and Schmidt decomposition
$\ket{\psi}=\sum_{i=1}^r\mu_i\ket{ii}$. It can be proved by
Lagrange multipliers that
\[
\max_{\{\varphi~s.t.~SR(\phi)\leq k\}}|\langle \psi | \varphi\rangle|^2=\sum_{i=1}^k \mu_i^2,
\]
where the maximum is taken with respect to states $\varphi$ which have at most Schmidt rank $k\leq r$, and
the Schmidt coefficients of $\psi$ are ordered as $\mu_i\geq\mu_{i+1}$. Thus, if
we want the witness $W_\epsilon(\psi)$ to be positive on states of
Schmidt rank $k$, we must have
\[
\epsilon\leq\frac{1-\sum_{i=1}^k\mu_i^2}{\sum_{i=1}^k\mu_i^2}.
\]
Notice that if $k=r$ we must put $\epsilon=0$. Correspondingly,
for $k< r$ (in the case $r\geq2$), let us define the Schmidt
number $k+1$ witness
\[
W_k(\psi)=\idmat-\frac{1}{\sum_{i=1}^k\mu_i^2}\pro{\psi}.
\]
Among $W_k(\psi)$'s, for fixed $\psi$, the witness able to detect
the largest number of entangled states is of course $W_1(\psi)$.
Yet, $W_1(\psi)$ cannot detect any PPT bound entangled state, as
we now prove.

Every pure state $\psi$ in $\mathbb{C}^d\otimes\mathbb{C}^d$ can
be written as $\ket{\psi}=\sqrt{d}(A\otimes\idmat)\ket{\Psi_d^+}$,
with $\Tr(A^\dagger A)=1$. The Schmidt coefficients of $\psi$ are
given by the singular values of $A$. In particular, for the
largest Schmidt coefficient we have
$\mu_1=\|A\|_\infty=\sqrt{\|AA^\dagger\|_\infty}$, where
$\|X\|_\infty$ is the operator norm of $X$. We have
$(\pro{\psi})^\Gamma=A\otimes\idmat V A^\dagger\otimes\idmat$,
with $V$ the swap operator. Notice that $V\leq\idmat$.

The witness $W_1(\psi)$, does not detect a state $\rho$ as
entangled if and only if $\bra{\psi}\rho\ket{\psi}\leq\mu_1^2$. If
$\rho=\rho_{AB}$ is PPT, $\tilde{\rho}=\rho^\Gamma$ is a
normalized state and
\begin{equation}
\begin{split}
\bra{\psi}\rho\ket{\psi}
&=\Tr(\rho^\Gamma(\pro{\psi})^\Gamma)\\
&=\Tr(\tilde{\rho}(A\otimes\idmat)V(A^\dagger\otimes\idmat))\\
&\leq\Tr(\tilde{\rho}(AA^\dagger\otimes\idmat))\\
&=\Tr(\tilde{\rho}_A AA^\dagger)\\
&\leq\max_{\phi}\bra{\phi}AA^\dagger\ket{\phi}\\
&= \mu_1^2,
\end{split}
\end{equation}
where $\tilde{\rho}_A=\Tr_B(\tilde{\rho})$. A straightforward
proof can also be obtained by considering that the reduction
criterion~\cite{reduction} is weaker than the PT criterion. Thus,
for every PPT state $\rho=\rho_{AB}$ we have
$\rho\leq\rho_A\otimes\idmat$. For the state $\psi$ under
consideration:
\[
\bra{\psi}\rho\ket{\psi}\leq\bra{\psi}\rho_A\otimes\idmat\ket{\psi}=\Tr(\rho_A AA^\dagger)\leq\mu_1^2.
\]

Witnesses $W_k(\psi)$, $N\geq2$ are even worse in detecting PPT
entangled states. Yet, for every $2\leq k\leq r-1$, and for any
state $\phi$ with Schmidt rank $2\leq l\leq k$, $P_\phi\otimes
W_k(\psi)$ is a nondecomposable entanglement witness.

\section{Conclusions}

We have been able to prove that any state $\rho_p(\{\psi^{(i)}\})$
is entangled as soon as $p>0$, for any set
$\{\psi_{(i)}\}_{i=1}^M$, when at least one state $\psi_{(i)}$ is
entangled, except in the case with only two pure states
$\psi^{(i)}$ with the same Schmidt coefficients. In the latter
case the state could be entangled as well, but an entanglement
witness different from \eqref{eq:structwitness} would be required
to prove it.

The structure of the states $\rho_p(\{\psi^{(i)}\})$ is very
simple: all the entanglement appears to be concentrated in an
eigenvector of the mixed state, while the separable part, in the
suitable region of parameters, plays the role of a ``cover'',
which prevents the detection by partial transposition (and hence
distillation). We remark the resemblance of the class of states
with isotropic states, most evident when considering the special
case of maximally entangled pure states
$\{\psi^{(i)}=\Psi^+_{d_i}\}$. Indeed, isotropic states were
involved in the analysis which signed the first appearance of the
class of states $\rho_p(\psi^{(1)},\psi^{(2)})$ in
literature~\cite{boundishizaka}.

Our analysis differs from the one appearing
in~\cite{boundishizaka} and~\cite{IPppt}, because we focus on the
properties of the states, rather than on what they allow to do,
and we construct witnesses to detect entanglement both in the
bipartite and multipartite settings. Moreover, we generalize the
states to the case of the possible choice of many pure states,
i.e. from the states $\rho_p(\psi^{(1)},\psi^{(2)})$ to the states
$\rho_p(\{\psi^{(i)}\})$. It is clear that the states
$\rho_p(\{\psi^{(i)}\})$ can be modified, both in their separable
and entangled parts, to provide larger classes of PPT entangled
states. Indeed, the key point toward the construction of PPT
entangled states, is that the separable part of
$\rho_p(\{\psi^{(i)}\})$ is not only positive semidefinite under
partial transposition, but strictly positive, so that slightly
changing it does not affect the positivity condition. For example,
using states similar to the ones studied in this paper, it is
possible to prove that many of the positive maps that were
conjectured to be nondecomposable in \cite{piani2006}, are
actually so~\cite{piani2006bis}. One remarkable property of the
structure \eqref{eq:genstates} is that states
$\rho_p(\{\psi^{(i)}\})$ are completely characterized by a set of
pure states $\{\psi^{(i)}\}$ and a mixing parameter $p$. Moreover,
they are separable for $p=0$ and entangled -- for almost any
choice of $\{\psi^{(i)}\}$ with at least one $\psi^{(i)}$
entangled -- for $p>0$. We hope that the variety of parameters at
disposal through the choice of the set of pure states
$\{\psi^{(i)}\}$ (and of the mixing parameter $p$), could lead to
the study of interesting cases/effects both in the bipartite and
in multipartite setting. Moreover, the simplicity of the
structure of this class of states suggests that they might be used
to offer the first experimental verification of the existence, and
properties, of bound entanglement. Preliminary studies of the
robustness of these states under the action of noise, for some
proper choice of the two pure states $\psi^{(1)}$ and
$\psi^{(2)}$, confirms the fact that experimental construction
of these states should be possible with current technology.
A detailed analysis of the noise tolerance of the states
$\rho_p$, together with a study of the possible experimental
realization of these states will be presented elsewhere.

Notice that all the results regarding witnesses can be directly
translated into results regarding maps through the
Choi-Jamio{\l}kowski isomorphism, so that in this paper we provide
many examples of nondecomposable maps, useful to detect the
entanglement of PPT states. Moreover, through the analysis of
states $\rho_p(\{\psi^{(i)}\})$ and of the corresponding
witnesses, we could provide a relationship among the properties of
$k$-positivity, complete positivity and decomposability: any map
that is $k$-positive, for $k\geq2$, but not completely positive,
can be extended to a nondecomposable map. Thus, it seems that
further analysis of the property of $k$-positivity could not only
be useful to study the Schmidt-number property of states,
following~\cite{Schmidtnumber}, but the entanglement property
itself.

We thank H.J. Briegel, O. Guehne, M. Horodecki, P. Horodecki, R.
Horodecki, A. Kossakowski and W. A. Majewski for discussions. We
gratefully acknowledge support by Austrian Science Foundation
(FWF), EU (RESQ (IST 2001 37559), OLAQUI, IP SCALA).


\begin{thebibliography}{40}
\bibitem[{wlo()}]{wlog}
\bibinfo{note}{The problem can be cast in the form \eqref{eq:system0} w.l.o.g.
  because there is no normalization constraint on $\ket{\alpha}$ and
  $\ket{\beta}$, i.e. on matrices $\alpha$ and $\beta$, and we can use this
  freedom to get rid of one proportionality constant in discussing the
  condition.}
\bibitem[{foo({\natexlab{a}})}]{footnote1}
\bibinfo{note}{Given a composite system $A_1A_2:B_1B_2$, the realignment
  operation can be split in realignement with respect to subsystems $1$ and $2$
  (exactly as partial transpostion with respect to $A$ can be split into
  partial transpostion with respect to $A_1$ and $A_2$):
  $R\big(\ket{i_{A_1}m_{A_2}j_{B_1}n_{B_2}}\bra{k_{A_1}p_{A_2}l_{B_1}q_{B_2}}
  \big)=R_1(\ket{i_{A_1}j_{B_1}}\bra{k_{A_1}l_{B_1}})\otimes
  R_2(\ket{m_{A_2}n_{B_2}}\bra{p_{A_2}q_{B_2}})=
  \ket{i_{A_1}m_{A_2}k_{B_1}p_{B_2}}\bra{j_{A_1}n_{A_2}l_{B_1}q_{B_2}}$.}
\bibitem[{foo({\natexlab{b}})}]{footnote3}
\bibinfo{note}{Note that, if $\tilde{W}_\epsilon(\psi^{(1)},\psi^{(2)})$ is a
  witness for a given $\epsilon$, then also
  ${W}_\epsilon(\psi^{(1)},\psi^{(2)})$ is a witness for the same value of
  $\epsilon$, while the converse is not true: consider for example the case the
  $\psi^{(i)}$'s have the same Schmidt rank but not the same Schmidt
  coefficients.}
\end{thebibliography}
\end{document}